\pdfoutput=1
\documentclass[conference]{IEEEtran}
\usepackage{epsfig}
\usepackage{fancyhdr}
\usepackage{plain}
\usepackage{epigraph}

\usepackage{url}
\usepackage{hyperref}
\usepackage{csquotes}
\usepackage{setspace}
\usepackage{amsmath, bm}
\usepackage{amssymb}
\usepackage{multirow}
\usepackage{color}

\usepackage[flushleft]{threeparttable}
\usepackage{verbatim}
\usepackage{tabularx}
\usepackage{array}
\usepackage{amsmath}
\usepackage{cases}
\usepackage{longtable}

\usepackage{threeparttablex}
\usepackage{longtable}
\usepackage{alltt}
\usepackage{upquote}
\usepackage{eurosym}

\usepackage{lmodern}
\usepackage{array}
\usepackage{mathabx}

\usepackage{epstopdf}
\usepackage{pdfpages}

\usepackage{flushend}

\begin{document}

\title{\emph{Desiree}: a Refinement Calculus for \\ Requirements Problems}

\begin{centering}
\author{\IEEEauthorblockN{Feng-Lin Li}
\IEEEauthorblockA{University of Trento\\
Trento, Italy\\
Email: fenglin.li@unitn.it} \\
\IEEEauthorblockN{Jennifer Horkoff}
\IEEEauthorblockA{City University\\
London, UK\\
Email: Horkoff@city.ac.uk}
\and
\IEEEauthorblockN{Alexander Borgida}
\IEEEauthorblockA{Rutgers University\\
New Brunswick, USA\\
Email: borgida@cs.rutgers.edu}\\
\IEEEauthorblockN{Lin Liu}
\IEEEauthorblockA{Tsinghua University\\
Beijing, China\\
Email: linliu@tsinghua.edu.cn}
\and
\IEEEauthorblockN{Giancarlo Guizzardi}
\IEEEauthorblockA{Federal University of Esp\'eito Santo\\
Vit\'oia, Brazil\\
Email: gguizzardi@inf.ufes.br}\\
\IEEEauthorblockN{John Mylopoulos}
\IEEEauthorblockA{University of Trento\\
Trento, Italy\\
Email: jm@disi.unitn.it}
}
\end{centering}
\maketitle

\begin{abstract}
The requirements elicited from stakeholders are typically informal, incomplete, ambiguous, and inconsistent. It is the task of Requirements Engineering to transform them into an eligible (formal, sufficiently complete, unambiguous, consistent, modifiable and traceable) requirements specification of functions and qualities that the system-to-be needs to operationalize. To address this requirements problem, we have proposed \emph{Desiree}, a requirements calculus for systematically transforming stakeholder requirements into an eligible specification. In this paper, we define the semantics of the concepts used to model requirements, and that of the operators used to refine and operationalize requirements. We present a graphical modeling tool that supports the entire framework, including the nine concepts, eight operators and the transformation methodology. We use a \emph{Meeting Scheduler} example to illustrate the kinds of reasoning tasks that we can perform based on the given semantics.
\end{abstract}

\IEEEpeerreviewmaketitle
\section{Introduction}

Upon elicitation, requirements are typically mere informal approximations of stakeholder needs that the system-to-be must fulfill. The key Requirements Engineering (RE) problem is to transform these requirements into a specification that describes formally and precisely the functions and qualities of the system-to-be. This problem has been elegantly characterized by Jackson and Zave~\cite{jackson_deriving_1995} as finding the specification \emph{S} that for certain domain assumptions \emph{DA} entails given requirements \emph{R}, and was formulated as $DA, S \models R$. Here domain assumptions circumscribe the domain where \emph{S} constitutes a solution for \emph{R}.

The RE problem is compounded by the very nature of the requirements elicited from stakeholders. They are often ambiguous, incomplete, unverifiable, conflicting, or just plain wrong~\cite{committee_ieee_1998}. In earlier studies on the PROMISE requirements dataset~\cite{menzies_promise_2012}, we found that 3.84\% of the 625 (functional and non-functional) requirements are ambiguous~\cite{li_stakeholder_2015}, 25.22\% of the 370 non-functional requirements (NFRs) are vague, and 15.17\% of the NFRs are potentially unattainable (e.g., they implicitly/explicitly use universals like ``\emph{any}'' as in ``any time'')~\cite{li_non-functional_2014}.

Our \emph{Desiree} framework tackles the RE problem in its full breadth and depth. In particular, it addresses issues of ambiguity (e.g., ``notify users with email'', where ``\emph{email}'' may be a means or an attribute of user), incompleteness (e.g., ``sort customers'', in ascending or descending order?), unattainability (e.g., ``the system shall remain operational at \emph{all} times'') and conflict (e.g., ``high comfort'' vs. ``low cost''). The \emph{Desiree} framework includes a modelling language for representing requirements (e.g., \emph{DA}, \emph{S}, and \emph{R}), as well as a set of refinement operators that support the incremental transformation of requirements into a formal, consistent specification. The refinement and operationalization operators strengthen or weaken requirements to transform what stakeholders say they want into a realizable specification of functions and qualities.

Refinement operators provide an elegant way for going from informal to formal, from inconsistent/unattainable to consistent, also from complex to simple. To support incremental refinement, we have proposed a description-based representation for requirements~\cite{li_stakeholder_2015}. Descriptions, inspired by AI frames and Description Logics (DL)~\cite{baader_description_2003}, have the general form ``\emph{Concept} $<$$slot_i$: $D_i$$>$'', where $D_i$ restricts $slot_i$; e.g., $R_1$ := ``{Backup} $<$actor: \{the\_system\}$>$ $<$object: Data$>$ $<$when: Weekday$>$''. This form offers intuitive ways to strengthen or weaken requirements. For instance, $R_1$ can be strengthened into ``Backup ... $<$object: Data$>$ $<$when: \{Mon, Wed, Fri\}$>$'', or weakened into ``Backup ... $<$object: Data$>$ $<$when: Weekday $\lor$ \{Sat\}$>$''. Slot-description (\emph{SlotD}) pairs ``$<slot: D>$'' allow nesting, hence ``$<$object: Data$>$'' can be strengthened to ``$<$object: Data $<$associated\_with: Student$>$$>$''. In general, a requirement can be strengthened by adding slot-description pair(s), or by strengthening a description. Weakening is the converse of strengthening. The notion of strengthening or weakening requirements maps elegantly into the notion of subsumption in DL and is supported by off-the-shelf reasoners for a subset of our language.

Our paper makes the following contributions:

\begin{itemize}
  \item Presents \emph{Desiree}, a holistic framework for transforming stakeholder requirements into an eligible specification.
  \item Formalizes the semantics of \emph{Desiree}, both the concepts and operators, by using Set theory.
  \item Introduces a GUI prototype tool that supports the entire framework, including the syntax, operators, a method for using the concepts and applying operators, and interrelations querying.
\end{itemize}


This work builds on our earlier studies~\cite{li_stakeholder_2015}\cite{li_non-functional_2014} that introduced our language for requirements and defined most of the operators. The new contributions of this work include the formal semantics for the language and the operators, the extension of some of the operators to address semantic problems of earlier versions, a GUI supporting tool and several kinds of reasoning tasks.



The rest of the paper is structured as follows. Section~\ref{cha:desiree_framework} introduces \emph{Desiree}, Section~\ref{cha:desiree_semantics} formalizes the semantics, Section~\ref{cha:tool} describes the supporting tool, and Section~\ref{cha:case_study} presents a \emph{Meeting Scheduler} case study. Section~\ref{cha:state_art} discusses related work, while Section~\ref{cha:conclusion} concludes, and sketches directions for further research.

\section{The \emph{Desiree} Framework}
\label{cha:desiree_framework}
In this section, we present the \emph{Desiree} framework, including a set of requirement concepts, a set of requirements operators, and a description-based syntax for representing these concepts and operators (we refer interested readers to Li~\cite{li_desiree_2016} for the systematic methodology for transforming stakeholder requirements into an eligible specification).

\subsection{Requirements Concepts}
\label{sec:concepts}
\begin{figure}[!htbp]
  \centering
  \vspace {-0.2 cm}
  \includegraphics[width=0.5\textwidth]{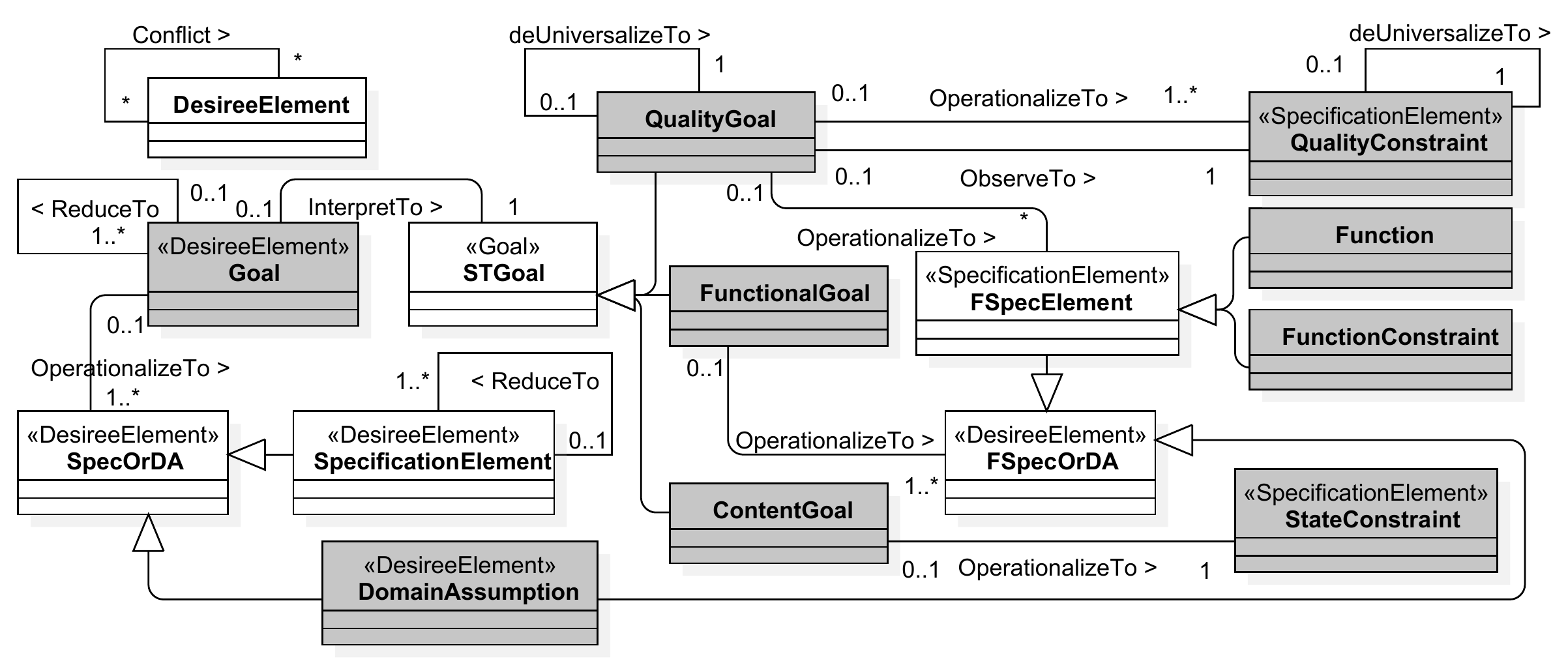}\\
  \vspace {-0.2 cm}
  \caption{The requirements ontology}\label{fig:desiree_concepts}
  \vspace {-0.5 cm}
\end{figure}

The core notions of \emph{Desiree} are shown in Fig.~\ref{fig:desiree_concepts}(shaded in the UML model). As in goal-oriented RE, we capture stakeholder requirements as goals. We have 3 sub-kinds of goals, 4 sub-kinds of specification elements (those with stereotype ``Specification Element''), and domain assumptions, all of which are subclasses of ``Desiree Element''. These concepts are derived from our ontological interpretation for requirements~\cite{li_non-functional_2014}\cite{li_stakeholder_2015} and our experiences on examining the large PROMISE requirements set~\cite{menzies_promise_2012}. We use examples from this set to illustrate each of these concepts and relations (see Li~\cite{li_desiree_2016} for the full syntax).

There are three points to be noted. First, `$>$' and `$<$' are used to indicate the reading directions of the relations. Second, the relations (except ``Conflict'') are derived from applications of requirements operators (to be discussed in Section~\ref{sec:opeartors}): if an operator is applied to an X object then the result will be $n..m$ Y objects; if that operator is not applied, then there is no relation (thus we have $0..1$ as the lower bound). Third, ``STGoal'', ``SpecOrDA'', ``FSpecOrDA'' and ``FSpecElement'' are artificial concepts, used to represent the union of their sub-classes, e.g., ``STGoal'' represents the union of ``FunctionalGoal'', ``QualityGoal'' and ``ContentGoal''. These classes are added to overcome the limitations of UML in representing inclusive (e.g., an operationalization of a functional goal) and exclusive (e.g., an interpretation of a goal) OR.

\vspace{6pt}
\noindent \emph{\textbf{Functional Goal, Function and Functional Constraint.}} A functional goal (\textbf{FG}) states a desired state, and is operationalized to one or more functions (\textbf{F}). For example, the goal ``student records be managed'' specifies the desired state ``managed''. We capture the intention of something to be in a certain state (situation) by using the symbol ``$:<$''. So this example is interpreted as a functional goal ``$FG_1$ := Student\_record $:<$ Managed'' (here ``Managed'' refers to an associated set of individuals that are in this specific state). This FG will be operationalized using functions such as ``add'', ``update'' and ``remove'' on student records.


As most perceived events in the Unified Foundational Ontology (UFO)~\cite{guizzardi_ontological_2005} are \emph{polygenic}, i.e., when an event is occurring, there are a number of dispositions of different participants being manifested at the same time, many pieces of information (e.g. restrictions over actor, object, and trigger) can be associated with the desired capability when specifying a function (\textbf{F}). For example, an execution of ``product search'' will involve participants like the system, a user, and product info. That is, ``\emph{the system shall allow users to search products}'' can be operationalized as a function ``$F_1$ := Search $<$subject: \{the\_system\}$>$$<$actor: User$>$$<$object: Product$>$''. Moreover, we can further add search parameters by adding a slot-description pair ``$<parameter: Product\_name>$''.

A functional constraint (\textbf{FC}) constrains the situation under which a function can be manifested. As above, we specify intended situations using ``$<s: D>$'' pairs and constrain a function or an entity involved in a function description to be in such a situation using ``$:<$''. For example, ``\emph{only managers are able to activate debit cards}'' can be captured as ``$FC_1$ := Active $<$object: Debit\_card$>$ $:<$ $<$actor: ONLY Manager$>$''.

\vspace{6pt}
\noindent \emph{\textbf{Quality Goal and Quality Constraint.}} We treat a quality as a mapping function that maps its subject to its value. Quality goal (\textbf{QG}) and quality constraint (\textbf{QC}) are requirements that require a quality to have value(s) in a desired quality region (QRG). In general, a QG/QC has the form ``Q (SubjT) :: QRG''. For instance, ``\emph{the file search function shall be fast}'' will be captured as a QG ``Processing\_time (File\_search) :: Fast''.

There are three points to be noted. First, QGs and QCs have the same syntax, but different kinds of QRGs: regions of QGs are vague (e.g., ``\emph{low}'') while those of QCs are often measurable (e.g., ``[0, 30 (\emph{Sec.})]'', {but see Example 5 in Section~\ref{sec:opeartors} for a more complex expression). Second, a quality name indicates a quality type, not a quality instance. By applying a quality type (e.g., ``Processing\_time'') to an individual subject \emph{x} of type \emph{SubjT} (e.g., a run of search, say \emph{search}\#1), we first get a particular quality \emph{q\#} (e.g., \emph{processing\_time}\#1), and then the associated quality value of \emph{q\#}. Third, when the subject is represented as individuals, we use curly brackets to indicate a set, e.g., ``\{the\_system\}''.

\vspace{6pt}
\noindent \emph{\textbf{Content Goal and State Constraint.}} A content goal (\textbf{CTG}) often specifies a set of properties of an entity in the real word, including both attributes and qualities, and these properties need to be represented by a system-to-be. To satisfy a CTG, a system needs to be in a certain state, which represents the desired world state. That is, concerned properties of real-world entities should be captured as data in the system. We use a state constraint (\textbf{SC}) to specify such desired system state.

For example, to satisfy the CTG ``\emph{A student shall have Id, name and GPA}'', the student record database table of the system must include three columns: Id, name and GPA. This example can be captured as a content goal ``$CTG_1$ := Student $:<$ $<$has\_id: ID$>$ $<$has\_name: Name$>$ $<$has\_gpa: GPA$>$'' and a state constraint, ``$SC_2$ := Student\_record :$<$ $<$ID: String$>$ $<$Name: String$>$ $<$GPA: Float$>$''.

\vspace{6pt}
\noindent \emph{\textbf{Domain Assumption.}} A domain assumption (\textbf{DA}) is an assumption about the operational environment of a system. For instance, ``\emph{the system will have a functioning power supply}'', which will be captured as ``$DA_{1}$ := \{the\_system\} :$<$ $<$has\_power: Power$>$'' using our language. Note that the syntax for DAs is similar to that for FCs. The difference is that an FC requires a subject to possess certain properties (e.g., in certain situations), while a DA assumes that a subject will have certain properties. In addition, DAs are also used to capture domain knowledge, e.g., ``\emph{Tomcat is a web server}'' will be captured as ``$DA_{2}$ := Tomcat $:<$ Web\_server''.

\subsection{Requirements Operators}
\label{sec:opeartors}
In this section, we introduce the set of requirements operators used for transforming requirements, which are inspired by traditional goal modeling techniques, and the syntactic form and semantics of our language. For example, since a QG/QC has the form ``Q (SubjT) :: QRG'', there can be different ways of refining it, based on whether Q, SubjT, or QRG is adjusted. In addition, since the semantics of such formulas have the form ``$\forall$ \emph{x}/\emph{SubjT}'', we also need to consider de-Universalizing them. Moreover, as qualities are measurable or observable properties of entities, we should also be able to add information about the observers who observe the quality.

In general, \emph{Desiree} includes two groups of operators (8 kinds in total): \emph{refinement} and \emph{operationalization}. An overview of these operators is shown in Table~\ref{tab:desiree_operators}, where ``\#'' means cardinality, ``\emph{m}'' and ``\emph{n}'' are positive integers (\emph{m} $\geq$ 0, \emph{n} $\geq$ 2). As shown, ``Reduce'', ``Interpret'', ``de-Universalize'', ``Scale'', ``Focus'' and ``Resolve'' are sub-kinds of refinement; ``Operationalize'' and ``Observe'' are sub-kinds of operationalization.


In the \emph{Desiree} framework, refinement operators are applied in the same category of elements: they refine goals to goals, or specification elements to specification elements. Operationalization operators map from goals to specification elements. Note that we do not support refinements from specifications to goals (i.e., requirements).

\begin{table}[!htbp]
  \vspace {-0.2 cm}
  \caption {An overview of the requirements operators }
  \label{tab:desiree_operators}
  \centering
  \scriptsize
  \setlength\tabcolsep{2pt}
  \begin{tabular}{|c|c|c|c|}
  \hline
  \multicolumn{2}{|c|}{\textbf{Requirements Operators}} & \textbf{\#InSet} & \textbf{\#OutSet} \\ \hline
  \multirow{6}{*}{\emph{Refinement}} & Reduce ($R_d$) & 1 & $1 ... m$ \\
  & Interpret (\emph{{I}}) & 1 & 1 \\
  & de-Universalize (\emph{{U}}) & 1 & 1 \\
  & Scale (\emph{{G}}) & 1 & 1 \\
  & Focus ($F_k$) & 1 & $1 ... m$ \\
  & Resolve ($R_s$) & $2 ... n$ & $0 ... m$ \\ \hline
  \multirow{2}{*}{\emph{Operationalization}}  & Operationalize ($O_p$) & 1 & $1...m$ \\
  & Observe ($O_b$) & 1 & 1 \\ \hline
  \end{tabular}
\end{table}

\textbf{\textbf{Reduce}} ($\bm{R_d}$). ``Reduce'' is used to refine a composite element (goal or specification element) to simple element(s), a high-level element to low-level element(s), or an under-specified element to sufficiently complete element(s). For instance, the composite goal $G_1$ ``\emph{collect real time traffic info}'' can be reduced to $G_2$ ``\emph{traffic info be collected}'' and $G_3$ ``\emph{collected traffic info be in real time}''.

The signature of operator ``${R_d}$'' is shown in Eq.~\ref{eq:reduce}, where $E'$ is a goal (e.g., goal, FG, QG, CTG) or a specification element (e.g., F, FC, QC, SC). It takes as input an element \emph{$E'$} and outputs a non-empty set (indicated by $\wp_1$, where $\wp$ represents power-set) of elements that are exactly of the same kind (with optional DAs). That is, we only allow reducing from goal to goal (not its sub-kind), FG to FG, F to F, etc; we also allow making explicit domain assumptions when applying the ``$R_d$'' operator. For example, when reducing $G_1$ ``\emph{pay for the book online}'' to $G_2$ ``\emph{pay with credit card}'', one needs to assume $DA_3$ ``\emph{having a credit card with enough credits}''. This refinement can be captured as ``$\bm{R_d}$ ($G_1$) = \{$G_2$, $DA_3$\}''.
\begin{equation}\label{eq:reduce}
  {\bm{R_d}: \; E' \rightarrow \bm{\wp_1} (E' \cup DA) }
\end{equation}

The ``$R_d$'' operator allows us to refine an element (a goal or a specification element) to several sub elements; hence it captures AND-refinement in traditional goal modeling techniques. To capture OR-refinement, we can apply the reduce operators several times, according to the different ways that a goal can be refined. For example, we will have two refinements ``$R_d$ ($G_1$) = \{$G_2$\}'' and ``$R_d$ ($G_1$) = \{$G_3$\}'' when reducing $G_1$ ``\emph{search products}'' to $G_2$ ``\emph{search by product name}'', and to $G_3$ ``\emph{search by product number}'', separately. As a result, ``$R_d$'' gives rise to a relation, not a (math) function.

\textbf{Interpret} ($\bm{I}$). The ``$\bm{I}$'' operator allows us to disambiguate a requirement by choosing the intended meaning, classify and encode a natural language requirement using our description-based syntax. For example, a goal $G_1$ ``\emph{notify users with email}'' can be interpreted as $FG_2$ ``User $:<$ Notified $<$means: Email$>$''. Note that $G_1$ is ambiguous since it has another interpretation $FG_3$ ``User $<$ has\_email: Email$>$ $:<$ Notified''. In such situation, analysts/engineers have to communicate with stakeholders in order to choose the intended interpretation.

We show the signature of ``$\bm{I}$'' as in Eq.~\ref{eq:interpret}, where $E$ is a \emph{Desiree} element (a goal, a specification element or a domain assumption) stated in natural language (NL), $E'$ is a structured \emph{Desiree} element, and is a subclass of $E$ or of the same type of $E$. Using this syntax, the ``notify user'' example will be written as ``$\bm{I}$ ($G_1$) = $FG_2$'' (suppose that $FG_2$ is the intended meaning).
\begin{equation}\label{eq:interpret}
  \bm{I}: \; E \rightarrow E'
\end{equation}

\textbf{Focus} ($\bm{F_k}$). The $\bm{F_k}$ operator is a special kind of ``Reduce'', and is used for refining a QGC (a QG or a QC) to sub-QGCs, following special hierarchies of its quality type or subject, e.g., dimension-of, part-of. For instance, for $QG_1$ ``Security (\{the\_system\}) :: Good'', the quality type ``Security'' can be focused to its sub-dimensions, e.g., ``Confidentiality'', the subject ``the system'' can be replaced by some of its parts, e.g., ``the data storage module''. The key point of applying $F_k$ is that if a quality goal $QG_1$ is focused to $QG_2$, then $QG_1$ would logically imply $QG_2$. For instance, if the system as a whole is secure, then its data storage module is secure, too.

The $\bm{F_k}$ operator has the signature as in Eq.~\ref{eq:focus}. In general, it replaces the quality type (resp. subject type) of a QGC with given Qs (resp., SubjTs), and returns new QGCs. Using this syntax, the focus of $QG_1$ to a sub-goal $QG_3$ ``Security (\{data\_storage\}) :: Good'' can be obtained as ``$\bm{F_k}$ ($QG_1$, \{data\_storage\}) = \{$QG_3$\}''.\\
\begin{equation}\label{eq:focus}
    \begin{aligned}
        \bm{F_k}: \; QGC \times \bm{\wp_1} (Q) \rightarrow \bm{\wp_1} (QGC) \\
        \bm{F_k}: \; QGC \times \bm{\wp_1}(SubjT) \rightarrow \bm{\wp_1} (QGC)
    \end{aligned}
\end{equation}

\textbf{Scale} ($\bm{G}$). In general, the $\bm{G}$ operator is used to enlarge the boundary of the quality region (QRG) of a QG/QC, in order to tolerate some deviations of quality values (i.e., relaxing the QG/QC to some degree). For instance, we can relax ``\emph{fast}'' to ``\emph{nearly fast}'', or ``\emph{in 30 seconds}'' to ``\emph{in 30 seconds with a scaling factor 1.2}''. The $\bm{G}$ operator can be also used to shrink the region of a QG/QC, strengthening the quality requirement. For example, we can replace a region ``\emph{good}'' with a sub-region ``\emph{very good}'', or more cleanly, replace ``[0, 30 (\emph{Sec.})]'' with ``[0, 20 (\emph{Sec.})]''. Based on the two cases, we specialize the ``\emph{scale}'' operator into ``\emph{scale down}'' ($\bm{G_d}$, relaxing through enlarging a QRG) and ``\emph{scale up}'' ($\bm{G_u}$, strengthening through shrinking a QRG).

The two scaling operators have the syntax as in Eq.~\ref{eq:scale}. They take as input a QG (resp. QC), a qualitative (resp. quantitative) factor and return another QG (resp. QC).
\begin{equation}\label{eq:scale}
    \begin{aligned}
        \bm{G}: QG \times QualitativeFactor \rightarrow QG \\
        \bm{G}: QC \times QuantitativeFactor \rightarrow QC	
    \end{aligned}
\end{equation}

Using this syntax, the relaxation of ``\emph{fast}'' to ``\emph{nearly fast}'' can be captured as in Example 1, and the enlarging of ``[0, 30 (Sec.)]'' to ``[0, 36 (Sec.)]'' through a pair of scaling factors ``(1, 1.2)'' can be written as in Example 2.

\begin{table}[!htbp]
    \centering
    \small
    \vspace{-0.2cm}
    \label{example:G_examples}
    \begin{tabular}{|l|}
        \hline
        {|\textbf{Example 1}|} \\ \hline
        $QG_{1-1}$ := Processing\_time (File\_search) :: Fast \\
        $QG_{1-2}$ := $\bm{G_d}$ ($QG_{1-1}$, Nearly) \\
        \hline
        {|\textbf{Example 2}|} \\ \hline
        $QC_{2-1}$ := Processing\_time(File\_search) :: [0, 30 (Sec.)] \\
        $QC_{2-2}$ := $\bm{G_d}$ ($QC_{2-1}$, (1, 1.2)) \\
        \hline
    \end{tabular}
    \vspace{-0.2cm}
\end{table}
Qualitative factors can be used to either strengthen (e.g., ``\emph{very}'') or weaken (e.g., ``\emph{nearly}'', ``\emph{almost}'') QGs, scaling their regions up and down, respectively. Quantitative factors are real numbers. We restrict ourselves to single-dimensional regions, which are intervals of the form $[Bound_{low} \; ... \; Bound_{high}]$. When enlarging a quality region, the scaling factor for $Bound_{low}$ shall be less than or equal to 1.0, and the factor for $Bound_{high}$ shall be greater than or equal to 1.0; when shrinking a region, opposite constraints must hold. Note that a region can be enlarged or shrunk, but not can be shifted. For example, we do not allow the change from ``[10, 20]'' to ``[15, 25]'', which is a shift. This is because we want to ensure the subsumption relation between regions when scaling them.

\textbf{de-Universalize} ($\bm{U}$). $\bm{U}$ applies to QGs and QCs to relax quality requirements, such that it is no longer expected to hold ``universally'', i.e., not to hold for 100\% of the individuals in a domain. For example, going from $QG_{3-1}$ ``(\emph{all}) \emph{file searches shall be fast}'' to $QG_{3-2}$ ``(\emph{at least}) \emph{80\% of the searches shall be fast}'' in Example 3.

Note that a QGC (QG/QC) description has a built-in slot ``\textbf{inheres\_in}'', relating a quality to the subject to which it applies. For example, the right-hand side (RHS) of $QG_{3-1}$, ``Processing\_time (File\_search)'', is actually ``Processing\_time $<$inheres\_in: File\_search$>$''. The syntax of $\bm{U}$ refers to a subset ``?X'' of the subject concept by pattern matching, using the SlotD ``$<$inheres\_in: ?X$>$''. Hence the above relaxation will be captured as in Example 3, where ``?X'' represents a sub-set of ``File\_search''.
\begin{table}[!htbp]
    \centering
    \small
    \vspace{-0.2cm}
    \begin{tabular}{|l|}
        \hline
        {|\textbf{Example 3}|} \\ \hline
        $QG_{3-1}$ := Processing\_time (File\_search) :: Fast \\
        $QG_{3-2}$ := $\bm{U}$ (?X, $QG_{3-1}$, $<$inheres\_in: ?X$>$, 80\%) \\
        \hline
    \end{tabular}
    \vspace{-0.2cm}
\end{table}

The general signature of U is given in Eq.~\ref{eq:deUniversalize}.
\begin{equation}\label{eq:deUniversalize}
        \bm{U}: varId \times QGC \times SlotD \times [0\%...100\%] \rightarrow QGC
\end{equation}

Sometimes we want to capture relaxations of requirements over requirements, for example, ``\emph{system functions shall be fast at 90\% of the time}'', relaxed to ``(\emph{at least}) \emph{80\% of the system functions shall be fast $($at least$)$ 90\% of the time}''. For this, we use nested $\bm{U}$s, as in Example 4 below. Here, ``?F'' is a sub-set of system functions (i.e., ``?F'' matches to ``System\_function''), and ``?Y'' is a sub-set of executions of a function in ``?F'' (i.e., ``?Y'' matches to the description ``Run $<$run\_of: ?F$>$'').

\begin{table}[!htbp]
    \centering
    \scriptsize
    \vspace{-0.2cm}
    \begin{tabular}{|l|}
        \hline
        {|\textbf{Example 4}|} \\ \hline
        $QG_{4-1}$ := Processing\_time (Run $<$run\_of: SysFunc$>$) :: Fast \\
        $QG_{4-2}$ := $\bm{U}$ (?F, $QG_{4-1}$, $<$inheres\_in: $<$run\_of:  ?F$>$$>$, 80\%) \\
        $QG_{4-3}$ := $\bm{U}$ (?Y, $QG_{4-2}$, $<$inheres\_in: ?Y$>$, 90\%) \\
        \hline
    \end{tabular}
    \vspace{-0.2cm}
\end{table}
$\bm{U}$ applies to only QGs and QCs. If one wants to specify the success rate of a function, one can first define a QC, e.g., ``Success (File\_search) :: True'', and then apply $\bm{U}$.

\textbf{Resolve} ($\bm{R_s}$). In practice, it could be the case that some requirements can stand by themselves, but will be conflicting when put together, since they cannot be satisfied simultaneously. Note that by conflict, we do not necessarily mean logical inconsistency, but can also be other kinds like normative conflict (e.g., a requirement could conflict with a regulation rule) or unfeasibility given the state of technology. For example, the goal $G_1$ ``\emph{use digital certificate}'' would conflict with $G_2$ ``\emph{good usability}'' in a mobile payment scenario. In \emph{Desiree}, we use a ``\emph{conflict}'' relation to capture this phenomenon, and denote it as ``\emph{Conflict} (\{$G_1$, $G_2$\})''.

The ``$\bm{R_s}$'' operator is introduced to deal with such conflicting requirements: it takes as input a set of conflicting requirements (more than one), while the output captures a set of compromise (conflicting-free) requirements, determined by the analyst. In the example above, we can replace $G_2$ by $G_2'$ ``acceptable usability'' or drop $G_1$.

The signature of $\bm{R_s}$ is shown in Eq.~\ref{eq:resolve}. Here we do not impose cardinality constraints on the output set, allowing stakeholders to totally drop the conflicting requirements when it is really necessary. Using this, we can write the resolution of this conflict as ``$\bm{R_s}$ (\{$G_1$, $G_2$\}) = \{$G_1$, $G_2'$\}'' or ``$\bm{R_s}$ (\{$G_1$, $G_2$\}) = \{$G_2$\}''.
\begin{equation}\label{eq:resolve}
    \bm{R_s}:  \bm{\wp_1} (E) \rightarrow \bm{\wp} (E)
\end{equation}

There are two points to be noted. First, analysts may declare known conflicts as DA axioms. For example, one can capture the conflict ``\emph{a user can not be both authorized and unauthorized}'' as ``$DA_1$ := Authorized ($\cap$) Unauthorized $:<$ Nothing''. Second, an application of the $\bm{R_s}$ operator will not physically delete any ``dropped'' requirement from a \emph{Desiree} model. For example, in the case of ``$\bm{R_s}$ (\{$G_1$, $G_2$\}) = \{$G_2$\}'', $G_1$ will still be kept in the model, but will not be considered during fulfillment reasoning (i.e., we do not consider if $G_1$ can be fulfilled, but do consider the remaining $G_2$). This is the same for $G_2$ in the case of ``$\bm{R_s}$ (\{$G_1$, $G_2$\}) = \{$G_1$, $G_2'$\}''.

\textbf{Operationalize} ($\bm{O_p}$). The $\bm{O_p}$ operator is used to operationalize goals into specification elements. In general, $\bm{O_p}$ takes as input one goal, and outputs one/more specification elements with optional domain assumptions. For instance, the operationalization of $FG_1$ ``Products :$<$ Paid'' as $F_2$ ``Pay $<$object: Product$>$ $<$means: Credit\_card$>$'' and $DA_3$ ``Credit\_card :$<$ Having\_enough\_credit'' will be written as ``$\bm{O_p}$ ($G_1$) = \{$F_2$, $DA_3$\}''.

The generalized syntax of $\bm{O_p}$ is shown in Eq.~\ref{eq:operationalize}.\\
\begin{equation}\label{eq:operationalize}
    \begin{aligned}
        \bm{O_p}: \; FG \rightarrow \bm{\wp} (F \cup FC \cup DA)	\\
        \bm{O_p}: \; QG \rightarrow \bm{\wp} (QC \cup F \cup FC \cup DA)	\\
        \bm{O_p}: \; CTG \rightarrow \bm{\wp} (SC \cup DA)	\\
        \bm{O_p}: \; Goal \rightarrow \bm{\wp} (DA)	\\
    \end{aligned}
\end{equation}

Note that one can use $\bm{O_p}$ to operationalize a QG as QC(s) to make is measurable, as Fs and/or FCs to make it implementable, or simply by connecting it to DA(s), assuming the QG to be true.

\textbf{Observe} ($\bm{O_b}$). The $\bm{O_b}$ operator is employed to specify the means, measurement instruments or human used to measure the satisfaction of QGs/QCs, as the value of slot ``\emph{\textbf{observed\_by}}''. For instance, we can evaluate ``\emph{be within 30 seconds}'' by assigning a stopwatch or assess a subjective QG ``\emph{the interface shall be simple}'' by asking observers. The $\bm{O_b}$ operator has the signature shown in Eq.~\ref{eq:observe}.
\begin{equation}\label{eq:observe}
        \bm{O_b}: \; (QG \cup QC) \times Observers \rightarrow QC
\end{equation}

Consider now the requirement ``(\emph{at least}) \emph{80\% of the surveyed users shall report the interface is simple}'', which operationalizes and relaxes the ``\emph{the interface shall be simple}''. The original goal will be expressed as $QG_{5-1}$ in Example 5. To capture the relaxation, we first use $\bm{O_b}$, asking a set of surveyed users to observe $QG_{5-1}$, and then use $\bm{U}$, to require (at least) 80\% of the users to agree that $QG_{5-1}$ hold. Here, the set variable ``?S'' represents a subset of surveyed users.

\begin{table}[!htbp]
    \centering
    \small
    \label{exmp:nested_u_examples}
    \begin{tabular}{|l|}
        \hline
        {|\textbf{Example 5}|} \\ \hline
        $QG_{5-1}$ := Style (\{the\_interface\}) :: Simple \\
        $QC_{5-2}$ := $\bm{O_b}$ ($QG_{5-1}$, Surveyed\_user) \\
        \; \; \; \; \; = $QG_{5-1}$ $<$observed\_by: Surveyed\_user$>$ \\
        $QC_{5-3}$ := $\bm{U}$ (?S, $QC_{5-2}$, $<$observed\_by: ?S$>$, 80\%)\\
        \hline
    \end{tabular}
\end{table}

\textbf{ReferTo}. We use the ``\emph{ReferTo}'' relation to capture the interrelations between Fs, FCs QGs, QCs, CTGs and SCs. In general, an F could refer to some CTGs or SCs, e.g., ``$F_1$ := Search $<$object: Product\_info$>$'' (``product info'' indicates a CTG); a QG/QC can take an F as its subject, containing its executions to be in certain time limitation, e.g., ``Processing\_time ($F_1$) :: [0, 30 (\emph{Sec.})]''; an FC could constrain an entity or a function involved in a function description, e.g., ``$F_1$ :$<$ $<$actor: ONLY Registered\_user$>$'' (only registered users can search); a CTG could refer to other SCs, e.g., when defining an attribute ``has\_product\_parameter'', we will need another SC ``Product\_parameter''.


In Li et al.~\cite{li_stakeholder_2015}, we have assessed the coverage of our requirements ontology by applying it to all the 625 requirements in the PROMISE dataset, evaluated the expressiveness of our description-based language by using it to rewrite all the 625 requirements in that dataset. Moreover, we have evaluated the effectiveness of the entire framework in Li et al.~\cite{li_engineering_2016} by conducting three controlled experiments, where upper-level software engineering students were invited to improve requirements quality by using our \emph{Desiree} approach and their ad-hoc approaches, respectively. The results provide strong evidence that with sufficient training (around two hours), \emph{Desiree} can indeed help people to identify and address more requirements issues (e.g., incompleteness, ambiguity and vagueness) when refining stakeholder requirements. 

\section{The Semantics of \emph{Desiree}}
\label{cha:desiree_semantics}
In this section, we provide the formal semantics of our language and operators. The former will be done using set theory, while the later will transform or relate formulas.





\subsection{The Semantics of the \emph{Desiree} Language}
\label{sec:semantics_language}

\textbf{Translation of \emph{Desiree} descriptions.} We show the syntax of \emph{Desiree} descriptions in the third column of Table~\ref{tab:semantics_descriptions} (see Li~\cite{li_desiree_2016} for the full syntax). In the forth column, we give the the translation of \emph{Desiree} descriptions to set-theoretic expressions, using the recursive function $\bm{T}$. When formalizing the semantics, we start from \emph{atomic} sets/types, for which elements we may do not know the structure. In our syntax, these correspond to \emph{\_Names} (e.g., \emph{ConceptNames}, \emph{RegionNames}) and \emph{RegionExpressions} (i.e., intervals and enumeration values); \emph{slots} are binary relations (the inverse of a slot $s$, is denoted as $s^{-1}$); \emph{ElementIdentifiers} are constants.

\begin{table}[!htbp]
  \centering
  \vspace {-0.2 cm}
  \begin{threeparttable}
  \caption {The semantics of \emph{Desiree} descriptions}
  \label{tab:semantics_descriptions}
  \scriptsize
  \setlength\tabcolsep{2pt}
  \begin{tabular}{|c|c|c|c|}
  \hline
  \textbf{Id} & & \textbf{Description} \emph{D}  & $\bm{T}$(\emph{D}) \\ \hline

  1 & \multirow{8}{*}{{\rotatebox[origin=c]{90}{\emph{\textbf{SlotD}}}}}
  & $<s: D>$
  & $\{x | \;  |s(x, \bm{T}(D))| = 1 \}$
  \\ \cline{1-1}\cline{3-4}

  2 &
  & $<s: \; \leq n \; D>$
  & $\{x | \; |s(x, \bm{T}(D))| \leq n \}$
  \\ \cline{1-1}\cline{3-4}

  3 &
  & $<s: \; \geq n \; D>$
  & $\{x |\; |s(x, \bm{T}(D))| \geq n \}$
  \\ \cline{1-1}\cline{3-4}

  4 &
  & $<s: n \; D>$
  & $\{x | \;  |s(x, \bm{T}(D))| = n \}$
  \\ \cline{1-1}\cline{3-4}

  5 &
  & $<s: {SOME} \; D>$
  & $\{x | \exists y . s(x,y) \land y \in \bm{T}(D)\}$
  \\ \cline{1-1}\cline{3-4}

  6 &
  & $<s: {ONLY} \; D>$
  & $\{x | \forall y. s(x, y) \rightarrow y \in \bm{T}(D)\}$
  \\ \cline{1-1}\cline{3-4}

  7 & &
  $SlotD_1 \; SlotD_2$
  & $\bm{T}(SlotD_1) \cap \bm{T} (SlotD_2)$
  \\ \cline{1-4} 

  8 & \multirow{8}{*}{{\rotatebox[origin=c]{90}{\emph{\textbf{Concept}}}}}
  & \emph{ConceptName}
  & \emph{ConceptName}
  \\ \cline{1-1}\cline{3-4}

  9 &
  & $SlotD$
  & $\bm{T}(SlotD)$
  \\ \cline{1-1}\cline{3-4}

  10 &
  & \{$ElemId_1$ ...\}
  & \{$ElemId_1$ ...\}
  \\ \cline{1-1}\cline{3-4}

  11 &
  & $D.s$
  & $\{x | \exists y. s^{-1}(x, y) \land y \in \bm{T}(D) \}$
  \\ \cline{1-1}\cline{3-4}

  12 &
  & $D_1 \; D_2$
  & $\bm{T}(D_1) \cap \bm{T} (D_2)$
  \\ \cline{1-1}\cline{3-4}

  13 &
  & $D_1 \vee D_2$
  & $\bm{T} (D_1) \cup \bm{T} (D_2)$
  \\ \cline{1-1}\cline{3-4}

  14 &
  & $D_1 - D_2$
  & $\{x | x \in \bm{T} (D_1) \land x \notin \bm{T} (D_2)\}$
  \\ \hline

  15 & \multirow{3}{*}{{\rotatebox[origin=c]{90}{\emph{\textbf{RgExpr}}}}}
  & $RegionName$
  & $RegionName$
  \\ \cline{1-1}\cline{3-4}

  16 &
  & $[AtomicVal_1, AtomicVal_2]$
  & $\{x|AtomicVal_1 \leq x \leq AtomicVal_2\}$
  \\ \cline{1-1}\cline{3-4}

  17 &
  & \{$AtomicVal_1$ ...\}
  & \{$AtomicVal_1$ ...\}
  \\ \hline
  \end{tabular}
  \begin{tablenotes}
      \scriptsize
      \item $s$: Slot; $D$: Description; Concept: \emph{Desiree} concept; RgExpr: region expression; $|M|$ denotes the cardinality of the set M, $s(x, M) := \{y | (x, y) \in s \land y \in M\}$, and the inverse of $s$, $s^{-1} := \{(x, y)| (y, x) \in s\}$.
    \end{tablenotes}
  \end{threeparttable}
  \vspace {-0.2 cm}
\end{table}

We start with seven basic rules for translating slot-description pairs (\emph{\textbf{SlotDs}}), the key constructor of our language (rule $1 \sim 7$). By default, a slot relates an individual to one instance that is of and only of type \emph{\textbf{D}}, a description that will be further defined (rule 1). Also, a \emph{SlotD} could have modifiers such as ``$\leq n$'',``$\geq n$'', ``$n$'' (``$n$'' is a positive integer), ``SOME'', or ``ONLY'' constraining its description (rule $2 \sim 6$). For instance, ``$<$register\_for:  $\geq$3 Class$>$'' represents a set of individuals that have registered for at least three classes. Two adjacent \emph{SlotDs} will be translated into their intersection (rule 7).

As shown in Table~\ref{tab:semantics_descriptions}, a description \emph{\textbf{D}} can be defined in various kinds of ways, such as an atomic concept name (denoting a set, e.g., ``Student\_record''), a slot-description (requiring its value to belong to a nested description, e.g., ``$<$register\_for: $\geq3$ Class$>$'', as we have shown before), or a set of individuals (e.g., the description ``Interoperable\_DBMS'' can be represented by a set of individuals ``\{MySQL, Oracle, MsSQL\}''); it can also be built using set constructors: intersection (represented by sequencing, e.g., ``Student $<$gender: Male$>$'', which captures ``male students''), union, and set difference (rule $12\sim14$). The expression \emph{D.s} (rule 11) is useful for describing the set of individuals related to elements of \emph{D} by \emph{s}. For instance, to capture ``\emph{the collected traffic info shall be in real time}'', we at first define a function ``$F_1$ := Collect $<$object: Traffic\_info$>$'', and then use ``$F_1$.object'', which is translated in to ``$\exists$ $object^{-1}$. $F_1$'' ($object^{-1}$ is the inverse of $object$), to refer to collected, instead of all, traffic info.

In addition, a description \emph{\textbf{D}} can be a region expression (rule $15 \sim 17$), which can be a region name (e.g., ``\emph{low}''), a mathematical region expression (e.g., ``$\geq$20'', as in ``Student $<$gender: Male$>$ $<$age: $\geq$20$>$'', which captures ``\emph{male students older than 20}''), or a set of enumeration values (e.g., ``Student $<$gender: Male$>$$<$age: \{20\}$>$'', which captures ``\emph{male students who are 20-years-old}'').

\emph{\textbf{Translation of requirements concepts.}} In general, stakeholder requirements are initially stated using natural language (NL). These early requirements are mere atomic descriptions~\footnote{In our framework, each kind of the requirements concepts, Goal, FG, CTG, QG, F, FC, QC, SC, or DA, can be initially stated in natural language and then be structured using the description-based syntax with the ``Interpret'' operator.}. The instances of an early requirement are all the corresponding problem situations (i.e., all the cases when the corresponding intention is instantiated). For example, for the requirement ``schedule a meeting'', the instances are all the specific requests for a meeting.


In our framework, stakeholder requirements are refined into \emph{Desiree} elements with a corresponding structured syntax. Especially, FGs, CTGs, FCs, SCs, and DAs are formulae of the form ``C $:<$ D'', where \emph{C} and \emph{D} are concepts defined using our language. The semantics of such formulae are closed formulae as in Eq.~\ref{eq:eq_semantics_subsumption}. For example, the semantics of the FC ``$FC_1$ := Data\_table $:<$ $<$accessed\_by: ONLY Manager$>$'' will then emerge as ``$\forall$\emph{x}/Data\_table $\forall$\emph{y} accessed\_by (\emph{x}, \emph{y}) $\rightarrow$ \emph{y} $\in$ Manager'', where ``$\forall x/C$'' is a shorthand for ``$\forall x.x \in C$''. In the rest of this section,  we focus on the semantics of the other three kinds of requirement concepts: Functions, QGs and QCs.\\
\begin{equation}\label{eq:eq_semantics_subsumption}
    \begin{aligned}
        \forall x.  x \in  \bm{T}(C) \rightarrow  x \in  \bm{T}(D) \\
    \end{aligned}
\end{equation}

\emph{\textbf{Translation of Functions.}} In \emph{Desiree}, a function description consists of a function name and a list of optional slot-description pairs (\emph{SlotDs}). Ontologically, it represents a set of manifestations (i.e., runs) of a capability. For example, the expression ``$F_1$ := Activate $<$actor: Manager$><$object: Debit\_card$>$'' says that each run of $F_1$ is an activation with a manager as its actor and a debit card as its object. Our description translation gives the semantics of $F_1$ as in Eq.~\ref{eq:eq_semantics_f_example_true}.\\
\begin{equation}\label{eq:eq_semantics_f_example_true}
    \begin{aligned}
    \forall x.F_1 (x)   & \rightarrow Activate (x)   \\
                        & \land |actor(x, Manager)| = 1 \\
                        & \land |object(x, Debit\_card)|=1\\
    \end{aligned}
\end{equation}

Note that we can not define the semantics of $F_1$ as in Eq.~\ref{eq:eq_semantics_f_example_false}, and restrict ``actor'' and ``object'' to be global functional slots (i.e., having only one instance of its description). This is because these slots may have multiple instances of their descriptions in other requirements in the same specification. For example, in the same specification, there could be another function ``$F_2$ := Activate $<$actor: Manager$>$$<$object: $\geq 1$ Debit\_card$>$'', where the ``object'' slot relates an execution of $F_2$ to a set of ($\geq 1$) debit cards. That is, $F_2$ allows a manager to activate debit cards in a batch mode.
\begin{equation}\label{eq:eq_semantics_f_example_false}
    \begin{aligned}
    \forall x.F_1 (x) & \rightarrow Activate (x)   \\
                      & \land \exists y.actor(x,y) \land \exists y.object(x,y)\\
    \end{aligned}
\end{equation}

In general, the semantics of a function specified as ``$F_e$ := \emph{FName} $<s: D>$'' can be generalized as in Eq.~\ref{eq:eq_semantics_f_general}.
\begin{equation}\label{eq:eq_semantics_f_general}
    \begin{aligned}
    \forall x/F_e \rightarrow FName(x) \land |s(x,\bm{T}(D))| = 1
    \end{aligned}
\end{equation}

We distinguish between a function individual $\{F\}$ and its manifestations $F$ since a function (capability) could have been implemented but not manifested if its activating situation does not hold. For example, a web site may have a keyword search capability, but will not be manifested if nobody use it. Further, the distinction between function (capability) and its manifestations allows us to specify requirements on the function (capability) itself. For example, we can not only require the ``schedule meetings'' function of a meeting scheduler (its manifestations, i.e., executions) to be fast, but also require the function (capability) itself to be easy to learn.

\emph{\textbf{Translation of QGs and QCs.}} A QGC (QG/QC) requires a quality to take its value in a desired region. For example, the quality constraint ``$QC_3$ := Processing\_time (File\_search) :: [0, 30 (Sec.)]'' requires each run of file search to take less than 30 seconds. Its semantics will be expressed by the formula in Eq.~\ref{eq:eq_semantics_qgc_example}, where ``\emph{inheres\_in}'' and ``\emph{has\_value\_in}'' are reserved binary predicates used to express the semantics of QGCs.
\begin{equation}\label{eq:eq_semantics_qgc_example}
    \small
    \begin{aligned}
      \forall s/File&\_search \; \forall q/Processing\_time \; inheres\_in( q, s) \\
                    & \rightarrow  has\_value\_in( q, region ( x, 0, 30, Sec.))
    \end{aligned}
\end{equation}

This can be generalized to QGCs, which have the syntax ``Q (SubjT) :: QRG'', by Eq.~\ref{eq:eq_semantics_qgc_general}.
\begin{equation}\label{eq:eq_semantics_qgc_general}
    \begin{aligned}
        \forall s/\bm{T}&(SubjT) \; \forall q/\bm{T}(Q) \; inheres\_in (q, s) \\
                        &\rightarrow  has\_value\_in(q, \bm{T}(QRG))
    \end{aligned}
\end{equation}

A QG has a qualitative region, e.g., ``\emph{fast}'', ``\emph{low}'', which is imprecise and is translated to a primitive concept, e.g., ``\emph{Fast}''. A QC has a quantitative region, which is a mathematically specified precise region. Since OWL2~\cite{group_owl_2009} supports one-dimensional primitive types only (i.e., it cannot represent points in $\mathbb{R}^3$), we restrict ourselves to single-dimensional regions, which are intervals of the form $[Bound_{low}, Bound_{high}]$ on the integer or decimal line.

\subsection{The Semantics of the Requirement Operators}
\label{sec:semantics_operator}


In this section, we provide the \emph{entailment} semantics for the requirement operators. When refining a goal $G_1$ to $G_2$:

\begin{enumerate}
  \item If each solution for $G_2$ is also a solution for $G_1$, then $G_2$ entails $G_1$ (denoted as $G_2$ $\models$ $G_1$), and this refinement is a \emph{strengthening};
  \item If each solution for $G_1$ is also a solution for $G_2$, then $G_1$ entails $G_2$ (denoted as $G_1$ $\models$ $G_2$), and this refinement is a \emph{weakening};
  \item In case $G_1$ and $G_2$ mutually entail each other (denoted as $G_1$ $\doteq$ $G_2$), we say that the two goals are equivalent and we term this refinement \emph{equating}.
\end{enumerate}




In \emph{Desiree}, an application of any requirements operator will be one of the three kinds of refinements: strengthening, weakening or equating. This entailment semantics is of importance to requirements refinement: we need to weaken a requirement if it is too strong (e.g., practically unsatisfiable, and conflicting) and constrain it if it is arbitrary (e.g., ambiguous, incomplete, and vague).
 
An overview of the entailment semantics of each operator is shown in Table~\ref{tab:semantics_operators}. 

\begin{table}[!htbp]
  \vspace {-0.2 cm}
  \caption {The entailment semantics of requirement operators}
  \label{tab:semantics_operators}
  \centering
  \small
  \setlength\tabcolsep{2pt}
  \begin{tabular}{|c|p{0.32\textwidth}|}
  \hline
  \textbf{Refinement} & \textbf{Operators} \\ \hline
  \multirow{5}{*}
  {Strengthening} & Interpret (Disambiguation) \\
	& Reduce (AND- and OR-refinement) \\
	& Reduce (Adding a SlotD, specializing the description of a slot) \\
	& Scale up (Shrinking QRG) \\
	& Operationalize (Goal as F, FC, QC, SC with optional DA) \\
	& Observe \\ \hline
  \multirow{5}{*}
  {Weakening} & de-Universalize \\
	& Scale down (Enlarging QRG) \\
	& Focus (Partial set of sub-elements) \\
    & Reduce (Removing a SlotD, generalizing the description of a slot) \\
	& Resolve \\
	& Operationalize (Goal as only DAs) \\ \hline
  \multirow{3}{*}
  {Equating} & Interpret (Encoding) \\
    & Reduce (Separating concern) \\
    & Focus (Full set of sub-elements) \\
    \hline
  \end{tabular}
  \vspace {-0.2 cm}
\end{table}

There are two points to be noted. First, a requirement operator that is ambiguous as to its strength status needs an additional argument to make the appropriate choice. We use `$\Dashv$' for a strengthening, `$\vDash$' for a weakening, and `$\doteq$' for an equating. For example, because an application of ``Reduce'' can be a strengthening, a weakening, or an equating, we could denote the reduce from $G_1$ to $G_2$ as $R_d$ ($G_1$, $\Dashv$) = \{$G_2$\} if the refinement is a strengthening, $R_d$ ($G_1$, $\vDash$) = \{$G_2$\} if it is a weakening, and $R_d$ ($G_1$, $\doteq$) = \{$G_2$\} if it is an equating. Second, we distinguish between ``Relators'', operators that assert relationships between existing elements (``Interpret'', ``Reduce'', ``Operationalize'', and ``Resolve''), and ``Constructors'', operators that construct in a precise way new elements from their arguments (``Focus'', ``Scale'', ``Observe'', ``de-Universalize''). We discuss each operator in detail in this sub-section.

\vspace{6pt}
\noindent {\textbf{Relators.}} Four out of the eight operators, namely ``Interpret'', ``Reduce'', ``Operationalize'', and ``Resolve'', are relators. These operators need to have their arguments ready, and then relate an input element to the corresponding output element(s) that are of concern. For example, given an ambiguous goal $G$, what the ``Interpret'' operator does is to choose one intended meaning from its multiple possible interpretations (i.e., the possible interpretations are already there, what we need to do is to discover them and choose the intended one from them). This is similar for ``Reduce'' or ``Operationalize'': choosing one kind of refinement/operationalization from multiple possible ones~\footnote{A \emph{Desiree} operator can be applied to the same input element more than one time. In an application, an input element can be refined/ operationalized to multiple sub-elements. This captures traditional AND. The multiple applications captures traditional OR.}.

\emph{\textbf{Reduce}} ($\bm{R_d}$). In general, an application of ``$\bm{R_d}$'' is a strengthening, but can also be an equating or a weakening. In general, traditional ``AND'' and ``OR'' refinements are strengthening. For example, when reducing a goal $G_1$ ``\emph{trip be scheduled}'' to $G_2$ ``\emph{hotel be booked}'' and $G_3$ ``\emph{airline ticket be booked}'', a solution that can satisfy both $G_2$ and $G_3$ can also satisfy $G_1$, but not the opposite since $G_1$ can be satisfied by other solutions (e.g., ``\emph{hostel be booked}'' and ``\emph{train ticket be booked}''). In this situation, we have $\bm{R_d}$ ($G_1$, `$\Dashv$') = \{$G_2$, $G_3$\}, which asserts $G_2, G_3 \models G_1$.

When reducing a complex requirement (a requirement with multiple concerns) to atomic ones (a requirement with a single concern), the refinement is an equating. For example, the reduction of $G_1$ ``\emph{the system shall collect real time traffic info}'' to $G_2$ ``\emph{traffic info be collected}'' and $G_3$ ``\emph{collected traffic info shall be in real time}''. In this case, having $\bm{R_d}$ ($G_1$, `$\doteq$') = \{$G_2$,$G_3$\} asserts $G_2$, $G_3$ $\doteq$ $G_1$.


When reducing a function description, an adding of a slot-description pair (\emph{SlotD}) or a specialization of the description of a slot is a strengthening, while a removing of a \emph{SlotD} or a generalization of the description of a slot is a weakening~\footnote{We do not allow strengthening a slot while weakening another in the same requirement description in a single refinement.}. For example, the refinement from ``$F_1$ := Book $<$object: Ticket$>$'' to ``$F_1'$ := Book $<$object: Airline\_ticket$>$'' is a strengthening since $F_1$ can be fulfilled by any solution that is able to book a ticket (e.g., airline ticket, bus ticket, train ticket, etc.), but $F_1'$ can only be fulfilled if a solution is able to book an airline ticket. That is, $F_1'$ has fewer solutions. This refinement can be captured as $\bm{R_d}$ ($F_1$, `$\Dashv$') = \{$F_1'$\}, which asserts $F_1' \models F_1$.



\emph{\textbf{Interpret}} ($\bm{I}$). In \emph{Desiree}, an interpretation of an ambiguous or under-specified requirement is a strengthening, and an encoding of a natural language requirement is an equating. This is because an ambiguous/under-specified requirement has more than one interpretation, hence possessing more solutions. For example, when an ambiguous goal $G_1$ ``\emph{notify users with email}'' is interpreted into $G_2$ ``\emph{notify users through email}'', each solution for $G_2$ could also be a solution for $G_1$, but not vice versa: $G_1$ has another interpretation $G_3$ ``\emph{notify users who have email}'', a solution for $G_3$ can fulfill $G_1$, but not $G_2$. In the case we simply encode a (natural language) requirement using our syntax, the encoding is an equating because the solution space does not change. Therefore, for two \emph{Desiree} elements $E_1$ and $E_2$, we have $E_2$ $\models$ $E_1$ if $\bm{I}$ ($E_1$, `$\Dashv$') = $E_2$, or $E_2$ $\doteq$ $E_1$ if $\bm{I}$ ($E_1$, `$\doteq$') = $E_2$.


\emph{\textbf{Operationalize}} ($\bm{O_p}$). Operationalization is similar to reduction ``$R_d$''. In \emph{Desiree}, the ``${O_p}$'' operator is overloaded in several ways. When operationalizing an FG to Function, FCs, DAs, or combinations thereof, or operationalizing a QG to QC(s) by making clear its vague region, to F and/or FCs to make it implementable, or operationalizing a CTG to SC(s), it is a strengthening. In case a goal is merely operationalized as DA(s), it is a weakening (a DA can be simply treated as true and can be fulfilled by any solution; that is, the solution space is enlarged to infinite when operationalizing a goal merely as DAs).

The $O_p$ operator has a similar semantics as ``Reduce'' in general, but is subtly different if instances of its output elements and instances of its input elements are not of the same type. For example, when an FG ``$FG_1$ := Meeting\_notification $:<$ Sent'' is operationalized as a function ``$F_2$ := Send $<$subject: \{the\_system\}$>$ $<$object: Meeting\_notification$>$'', we will have ``$F_2$.effect $\models$ $FG_1$'', where ``$F_2$.effect'' represent the situation brought about by the execution of $F_2$ (i.e., a message is sent). Note that here we can not use ``$F_2 \models FG_1$'' because the set of instances of $F_2$ and that of $FG_1$ are of different types: instances of $F_2$ are a set of its executions while instances of $FG_1$ are a set of problem situations, or, alternatively, instantiations of the corresponding intention (the intention of a message being sent, in this case). If $FG_1$ is operationalized to a set, say $F_2$ and $F_3$, we will have $F_2$.effect, $F_3$.effect $\models$ $FG_1$.


\emph{\textbf{Resolve}} ($\bm{R_s}$). An application of ``$\bm{R_s}$'' to a set of conflicting requirements $S_i$ will produce a set of conflict-free requirements $S_o$. Being a weakening, such a refinement will be denoted as ``$\bm{R_s}$($S_i$, `$\vDash$') = $S_o$''. A conflict among a set of requirements means that they cannot be satisfied simultaneously; that is, there is an empty set of solutions for $S_i$. Once resolved, then there should be some possible solutions; that is, there is a non-empty set of solutions for $S_o$. Hence this adds the meta-statement $S_i \models S_o$.


\vspace{6pt}
\noindent \textbf{Constructors.} The remaining four operators, namely ``Focus'', ``Scale'', ``Observe'', ``de-Universalize'', are constructors. These operators take as input \emph{Desiree} elements and necessary arguments, and create/construct new elements, for which entailment can be computed by logic.

\emph{\textbf{Focus}} ($\bm{F_k}$). The ${F_k}$ operator narrows the scope of a quality or subject by following certain hierarchies, e.g., ``dimension-of'' ``part-of'', and makes a QGC easier to fulfill. Its application leads to a weakening (e.g., focusing ``security'' to a sub-set of its dimensions, say ``confidentiality'', ``the system'' to some of its sub-parts ``interface''), but sometimes an equating (e.g., focusing ``security'' to the full set of its sub-dimensions, ``confidentiality'', ``integrity'' and ``availability''). In the former case, where $\bm{F_k}$ ($QGC_1$, $Qs$/$SubjTs$, `$\vDash$') = $QGC_{partial}$ ($QGC_{partial}$ is a sub-set of the potential resultant QGCs), we have $QGC_1$ $\models$ $QGC_{partial}$; in the latter case, where $\bm{F_k}$ ($QGC_1$, $Qs$/$SubjTs$, `$\doteq$') = $QGC_{full}$ ($QGC_{full}$ is the full-set of the the potential resultant QGCs), we have $QGC_1$ $\doteq$ $QGC_{full}$, i.e., $QGC_1$ $\models$ $QGC_{full}$ and $QGC_{full}$ $\models$ $QGC_1$.

\emph{\textbf{Scale}} (\emph{\textbf{G}}). The \emph{G} operator is used to enlarge or shrink quality regions. When enlarging a region of $QGC_1$ to $QGC_2$, e.g., scale ``\emph{fast}'' to ``\emph{nearly fast}'' or ``[0, 30 (\emph{Sec.})]'' to ``[0, 40 (\emph{Sec.})]'', logic gives $QGC_1 \models QGC_2$ (i.e., it is a weakening). When shrinking a region of $QGC_1$ to $QGC_2$, e.g., strengthen ``\emph{fast}'' to ``\emph{very fast}'' or ``[0, 30 (\emph{Sec.})]'' to ``[0, 20 (\emph{Sec.})]'', logic implies $QGC_2 \models QGC_1$ (i.e., it is a strengthening).


\emph{\textbf{de-Universalize}} ($\bm{U}$). The $U$ operator is used for weakening. For example, instead of requiring all the runs of file search to take less than 30 seconds, we can relax it to 80\% of the runs to be so by applying $U$. Formally, the semantics of $QG_{5-2}$ ``$\bm{U}$ (?X, $QG_{5-1}$, $<$inheres\_in: ?X$>$, 80\%)'', derived from $QG_{5-1}$ ``Processing\_time (File\_search) :: Fast'', is expressed as in Eq.~\ref{eq:eq_semantics_U_example}.
\begin{equation}\label{eq:eq_semantics_U_example}
    \begin{aligned}
        \exists & ?X/\bm{\wp}(File\_search) (|?X|/|File\_search| > 0.8 \\
                  & \land \forall s/?X \; \forall q/Processing\_time \; inheres\_in (q, s) \\
                  &  \rightarrow has\_value\_in(q,  Fast))
    \end{aligned}
\end{equation}


In general, the semantics of a QGC that is applied with $U$ (UQGC for short), of the form ``$\bm{U}$ (?X, Q (SubjT) :: QRG, $<$inheres\_in: ?X$>$, Pct)'', is given in Eq.~\ref{eq:eq_semantics_U}, where a percentage ``Pct'' indicates a region [Pct, 100\%]. We can see that if $QGC_1$ is relaxed to $QGC_2$ by $U$, logic gives $QGC_1$ $\models$ $QGC_2$. \\
\begin{equation}\label{eq:eq_semantics_U}
    \begin{aligned}
    \exists ?X/ & \bm{\wp}(\bm{T}(SubjT)) \; (|?X|/|\bm{T}(SubjT)| > Pct \\
                        & \land \forall s/?X \; \forall q/\bm{T}(Q) \; inheres\_in (q, s) \\
                        & \rightarrow has\_value\_in (q,  \bm{T}(QRG)))	
    \end{aligned}
\end{equation}

In addition, the $\bm{U}$ operator would have the property as shown in Eq.~\ref{eq_translation_uqgc_property}, where $0\% \leq Pct_2 < Pct_1 \leq 100\%$.
\begin{equation}\label{eq_translation_uqgc_property}
    \begin{aligned}
    \bm{U} (?X, QGC, <inheres\_in: \; ?X>, Pct_1) \models \\  \bm{U} (?X, QGC, <inheres\_in: \; ?X>, Pct_2)
    \end{aligned}
\end{equation}

In the case U is nested as in Example 5 (Section~\ref{sec:opeartors}), we refer interested readers to Li~\cite{li_desiree_2016} for the semantics.

\emph{\textbf{Observe}} ($\bm{O_b}$). An application of ``Observe'' to a QGC will append a \emph{SlotD} ``$<$observed\_by: Observer$>$'' to the QGC, hence strengthen it. That is, we will have $QGC_2 \models QGC_1$ when applying $O_b$ to $QGC_1$: $\bm{O_b}(QGC_1, Observer)$ = $QGC_2$. For example, by applying $O_b$, $QG_{7-1}$ ``Style (\{the\_interface\}) :: Simple'' becomes ``$QC_{7-2}$ := $QG_{7-1}$ $<$observed\_by: Surveyed\_user$>$'', the semantics of which is shown in Eq.~\ref{eq:eq_semantics_Ob_example}.\\
\begin{equation}\label{eq:eq_semantics_Ob_example}
    \begin{aligned}
         \forall o&/Surveyed\_user \; \forall s/\{the\_interface\} \; \forall q/Style \\
                                & (inheres\_in (q, s) \rightarrow observed\_by (q, o)) \\
                                & \rightarrow has\_value\_in (q, Simple)
    \end{aligned}
\end{equation}

To obtain the general semantics of $O_b$, we only need to replace the specific quality type (e.g., ``Style''), subject type (e.g., ``{the\_interface}''), quality region (e.g., ``Simple''), and observer (e.g., ``surveyed\_user'') with general \emph{Q}, \emph{SubjT}, \emph{QRG} and \emph{Observers}, as in Eq.~\ref{eq:eq_semantics_Ob_general}.\\
\begin{equation}\label{eq:eq_semantics_Ob_general}
    \begin{aligned}
        \forall o&/\bm{T}(Observer) \; \forall s/\bm{T}(SubjT) \; \forall q/\bm{T}(Q) \\
                                & (inheres\_in (q, s) \rightarrow observed\_by (q, o)) \\
                                & \rightarrow has\_value\_in (q, \bm{T}(QRG))
    \end{aligned}
\end{equation}

In the above example, $QC_{7-2}$ is hard to satisfy since it requires all the surveyed users to agree that the interface is simple, and is often relaxed using $U$. For instance, we could have ``$QC_{7-3}$ := $\bm{U}$ (?O, $QC_{7-2}$, $<$observed\_by: ?O$>$, 80\%)'', which requires only 80\% of the users to agree. We show the semantics of $QC_{7-3}$ in Eq.~\ref{eq:eq_semantics_Ob_U_example}.
\begin{equation}\label{eq:eq_semantics_Ob_U_example}
    \small
    \begin{aligned}
         \exists ?O&/\bm{\wp}(Surveyed\_user).[|?O|/|Surveyed\_user|> 0.8 \\
                                & \land \forall o/?O \; \forall s/\{the\_interface\} \; \forall q/Style \\
                                & (inheres\_in (q, s) \rightarrow observed\_by (q, o)) \\
                                & \rightarrow has\_value\_in (q, Simple) ]
    \end{aligned}
\end{equation}

Due to space limitation, we do not consider the ``\emph{fulfillment}'' semantics for each operator (i.e., the propagation of fulfillment from the output elements of an operator to its input) in this paper, and refer interested readers to Li~\cite{li_desiree_2016}, Horkoff et al.~\cite{horkoff_making_2012} for details.

\section{The \emph{Desiree} Tool}
\label{cha:tool}

In this section, we present a prototype tool that is developed in support of our \textit{Desiree} framework. As shown in Fig.~\ref{fig:tool_desiree_usage}, the \emph{Desiree} tool consists of three key components: (1) a textual editor, which allows analysts/engineers to write requirement using our syntax; (2) a graphical editor, which allows users to draw requirements models through a graphic user interface; (3) a reasoning component, which translates requirements (texts or models) to OWL2 ontologies, and make use of existing reasoners (e.g., Hermit~\cite{shearer_hermit:_2008}) to perform reasoning tasks.

\begin{figure}[!htbp]
  \centering
  \vspace {-0.3 cm}
  \includegraphics[width=0.5\textwidth]{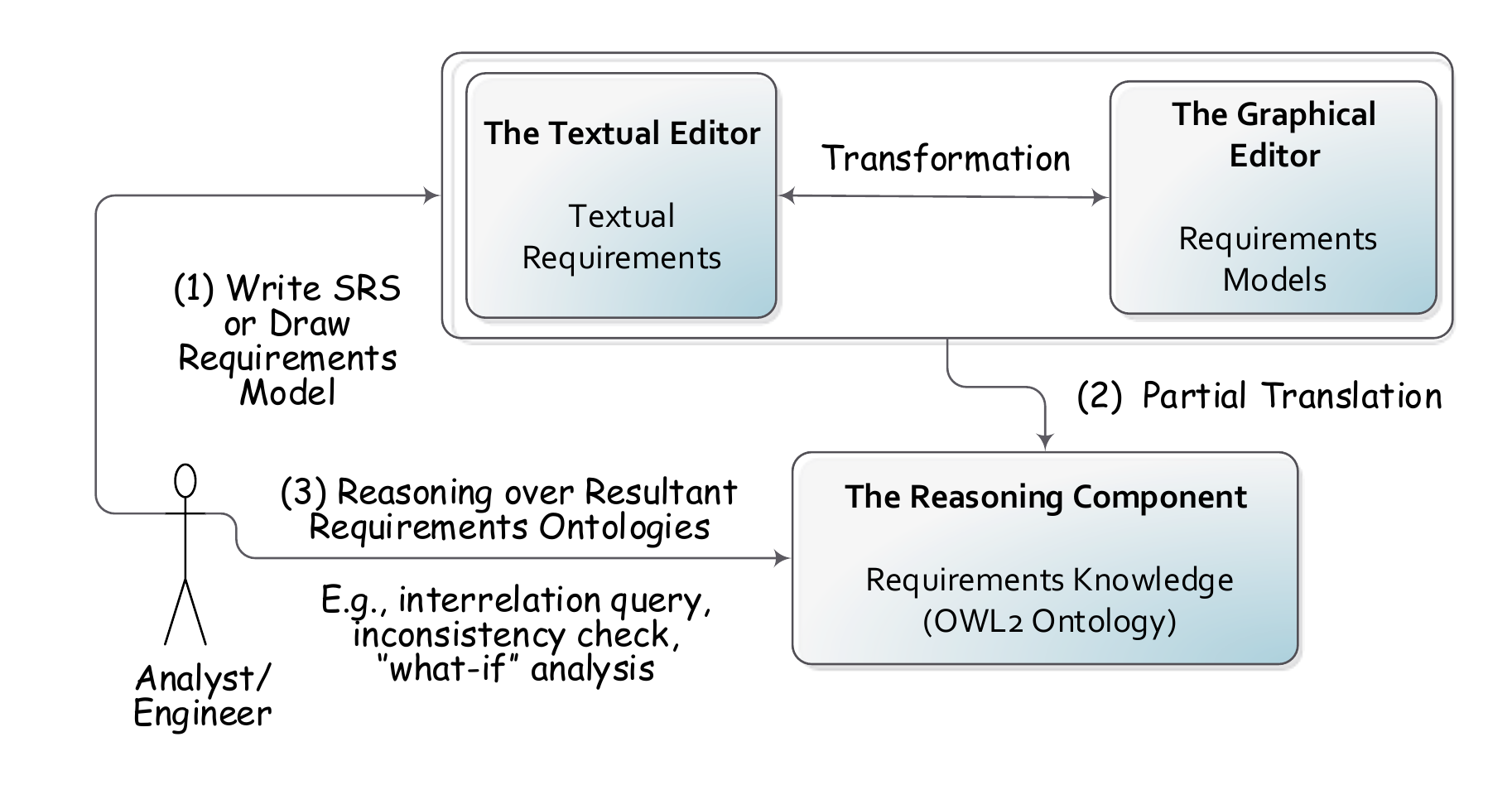}\\
  \vspace {-0.3 cm}
  \caption{The usage of the \emph{Desiree} tool}\label{fig:tool_desiree_usage}
  \vspace {-0.2 cm}
\end{figure}

One of the key features of the \emph{Desiree} tool is its strong support for scalability. This support is two-fold: (1) we allow analysts/engineers to write textual requirements specifications, and automatically create graphical models from textual specifications with built-in automatic layout; (2) we allow analysts/engineers to dynamically create user-defined views for large models, with each view showing a sub-part of the model.

\textbf{The Textual Editor}. Our textual editor offers several useful features: (1) syntax coloring, highlighting keywords in our syntax (e.g., the name of requirement concepts and operators will be colored as red); (2) content assistance, providing users with syntactic hints for our description-based syntax; (3) error checking, reporting syntax errors in requirements on the fly.

\textbf{The Graphical Editor}. Our graphical modeling tool provides several practical features: (1) it is able to transform textual requirements written in our syntax to graphical models with automatic layout, decreasing the efforts of drawing large models; (2) it maintains a central repository for each requirement model, and allows analysts/engineers to dynamically create views as needed (e.g., one can create a view for an important quality and its refinement), aiming to address the scalability of large models; (3) it allows analysts/engineers to filter models by choosing specific requirements concepts or refinements/operationalizations (e.g., one can choose to see only the reduce refinements in a user-defined view).



\textbf{The Reasoning Component}. This module includes two parts: (1) a parser that is built on OWL API and translates requirements (texts or models) specified using our language to OWL2 ontologies; (2) a reasoning part that makes use of an existing reasoner, i.e., Hermit~\cite{shearer_hermit:_2008}, to perform reasoning tasks. The translation to OWL2 (see Li~\cite{li_desiree_2016} for the details about translation) captures part of the semantics of our language (as the expressiveness of DL is much weaker, e.g., the expressions of nested \textbf{U} cannot be supported), but this translation still allows us to do some interesting reasoning such as interrelations query and inconsistency check (see the case study in Section~\ref{sec:case_study} for an illustration).

We show an overview of the tool in Fig.~\ref{fig:tool_graphical_editor}. Note that a requirements model in \emph{Desiree} can be consisting of multiple modeling views (canvas pages), e.g., the model in Fig.~\ref{fig:tool_graphical_editor} consists of two modeling views, ``Page-0'' and ``Page-1''. In addition, the global outline differs from a local outline in that the former sketches all the elements in a model (i.e., all the elements in its constituting canvas pages) while the latter outlines only the element in a specific canvas page.

\begin{figure}[!htbp]
  \centering
  \includegraphics[width=0.48\textwidth]{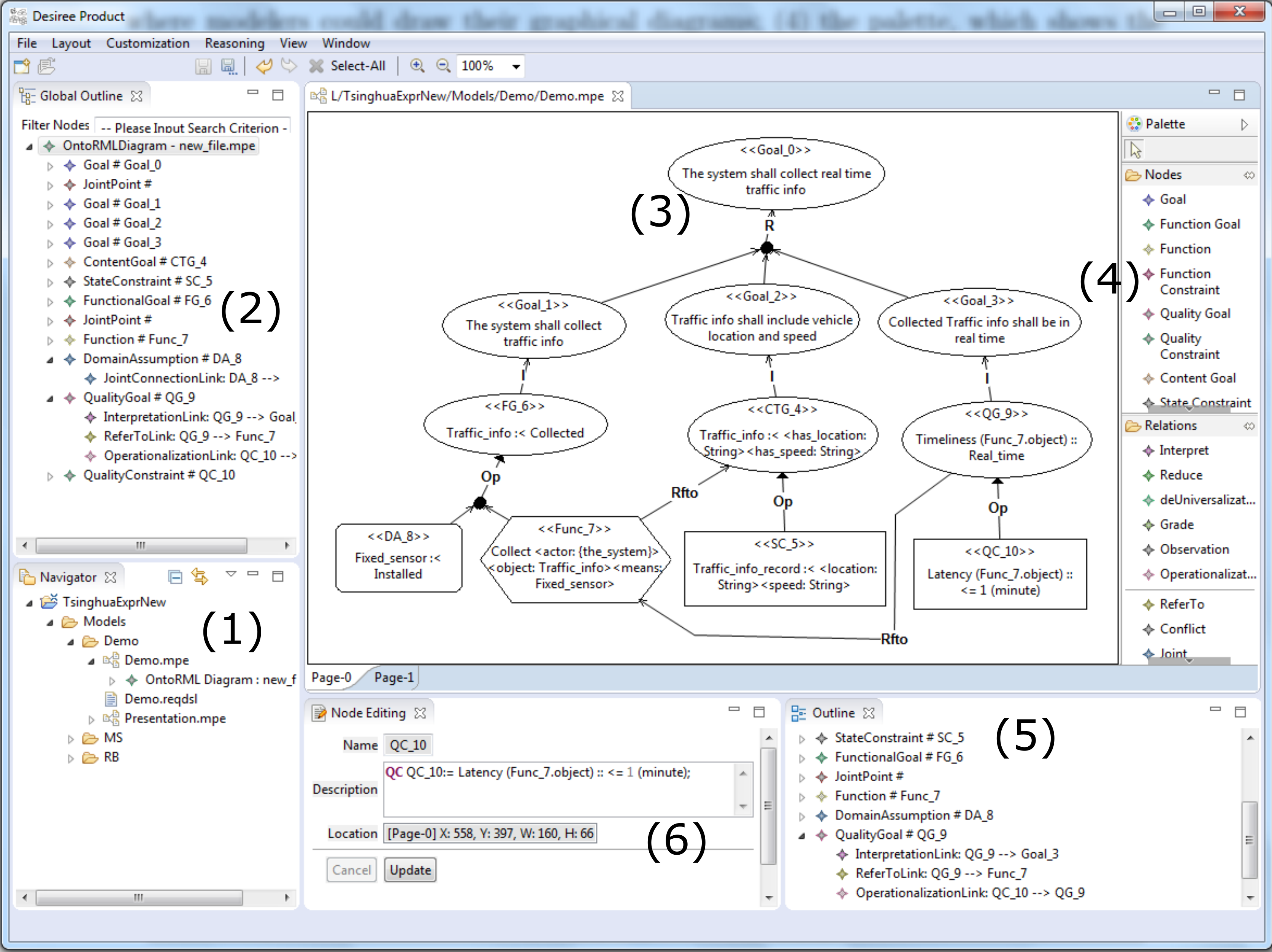}\\
  \vspace {-0.1 cm}
  \caption{An overview of the graphical editor}\label{fig:tool_graphical_editor}
  \vspace {-0.4 cm}
\end{figure}

\section{A Meeting Schedule Case Study}
\label{cha:case_study}

We performed a case study on the \emph{Meeting Scheduler} (MS) project, chosen from the PROMISE dataset~\cite{menzies_promise_2012}, to illustrate how our \emph{Desiree} framework can be applied to realistic requirements.

The MS project has 74 requirements (27 FRs and 47 NFRs). Functionally, the meeting scheduler is required to create meetings, send meeting invitations, book meeting rooms, and so on. The non-functional requirements cover different aspects of the system, such as ``Usability'', ``Configurability'', ``Look and feel'', ``Inter-operability'', ``Security'', and ``Maintainability''.

We first classified the 74 requirements according to our requirement ontology, and then encoded them by using our description-based syntax. During the process, we identified several kinds of requirements issues, such as ambiguous, incomplete, unverifiable, and unsatisfiable. We resolved these issues using the set of our provided operators (e.g., ``Interpret'', ``Reduce'', ``Operationalize'', ``Observe'', and ``de-Universalize'') to make them unambiguous, sufficiently complete, verifiable, and practically satisfiable.


We finally obtained a specification, which consists of 58 functions, 54 QCs, 10 FCs, 8 SCs and 13 DAs (143 elements in total). We kept the requirements, specification, and the derivation process (refinements and operationalizations) in a \emph{Desiree} model, and translated the model to an OWL2 ontology using the prototype tool.



The major benefit of such a translation is the convenience of obtaining an overview of concerns (e.g., functions, qualities and entities) and interrelations query: we are able to ask a list of questions as shown in Table~\ref{tab:ms_query} (technically, these questions will be translated into DL queries). For instance, we can ask ``$<$inheres\_in: \{the\_product\}$>$'' (an instantiation of Q2) to retrieve the set of qualities that inhere in ``the product''. Note that these questions are not exhaustive. If desired, we can ask more complex questions like ``what functions are required to finish within 5 sec.?'' in the form of ``$<$has\_quality: Processing\_time $<$has\_value\_in: $\leq$ 5 (Sec.)$>>$''.

\begin{table*}[!htbp]
  \vspace{-0.2cm}
  \caption {Example queries over the \emph{Meeting Scheduler} requirements specification }
  \label{tab:ms_query}
  \small
  \centering
  \begin{tabular}{|c|c|c|}
  \hline
  \textbf{ID} & \textbf{Concerned Questions } & \textbf{Our Syntax} \\ \hline
  Q1 & What kinds of subjects does a quality refer to?	& $<$has\_quality: QualityName$>$ \\ \hline
  Q2 & What qualities are of concern for a subject?	& $<$inheres\_in: SubjT$>$ \\ \hline
  Q3 & Who performs the function? &	$<$is\_actor\_of: F$>$ \\ \hline
  Q4 & What is the function operating on? & $<$is\_object\_of: F$>$ \\ \hline
  Q5 & What are the functions that an object is involved in? & $<$object: SubjT$>$ \\ \hline
  \end{tabular}
  \vspace{-0.2cm}
\end{table*}


A second benefit of such a translation is the identification of (some) inconsistencies. For example, the term ``user'' in ``\emph{users shall be able to register within 2 minutes}'' refers to a person in the real world, but a symbolic or representational entity in the system in ``\emph{managers shall be able add users into the system}''. To detect such inconsistencies, we add two axioms, ``System\_function $:<$ $<$object: ONLY Information\_entity$>$'' and ``Information\_entity ($\land$) Real\_world\_entity $:<$ Nothing'', constraining the object of a system function to be only information entities, which are disjoint with real-world entities. When specifying the function ``Register $<$actor: User$>$ $<$target: \{the\_system\}$>$'', we will add a DA axiom ``User $:<$ Real\_world\_entity''. If we have another function ``Add $<$actor: Manager$>$$<$object: User$>$$<$target: \{the\_system\}$>$'', and all these descriptions are translated into DL formulae, a DL reasoner is able detect an inconsistency therein: the term ``User'' cannot be used to refer to both a real-world entity and an information entity, which are disjoint classes, at the same time. Moreover, the identification of such inconsistencies helps us to discover implicit or missing requirements, e.g., there is an implicit content requirement about user profile in this example. In this case study, we identified 3 such inconsistencies, the other two are ``\emph{meeting room}'' and ``\emph{room equipment}'' (being a real-world entity vs. being an information entity).


Another benefit, which is closely related to the interrelation management, is the support for impact analysis when changes occur. For instance, for security reasons, a stakeholder may require their email to be invisible to others. Given this new requirement, we can at first find out the functions that are related to emails through query (suppose that we have ``$F_x$ := Send $<$object: Meeting\_invitation$>$ $<$means: Email$>$'' through the query ``$<$means: Email$>$''), and then add the new requirement according to some mechanisms (e.g., reduce $F_x$  to $F_x'$ := Send $<$object: Meeting\_invitation$>$ $<$means: Email $<$kind: BCC$>>$'', making ``\emph{blind carbon copy} (\emph{bcc})'' a nested \emph{SlotD} that modifies ``Email'', the means of $F_x$). Next, we need to evaluate how this change would impact other elements (e.g., the influence on the inhering performance qualities of $F_x$). This interesting topic will be explored in the next steps of our work. 
\section{Related Work}
\label{cha:state_art}

\subsection{Requirements Ontology}
\label{sec:art_fr_nfr}


An initial conceptualization for RE was offered by Jackson and Zave~\cite{jackson_deriving_1995} nearly two decades ago, founded on three basic concepts: \textit{requirement}, \textit{specification}, and \textit{domain assumption}. This initial characterization was extended by the Core Ontology for RE (aka CORE)~\cite{juretaa_core_2009} that differentiate non-functional requirements (NFRs) from functional requirements (FRs) using \textit{qualities} introduced in the DOLCE foundational ontology~\cite{masolo_ontology_2003}. In our experience, one of the deficiencies of CORE is that it can not classify requirements that refer to both qualities and functions.

Also, ontologies of specific domains, for which requirements are desired, have been employed in RE mainly for activities~\cite{kaiya_using_2006} or processes~\cite{falbo_evolving_2008}. These efforts, however, are not proposals for an ontological analysis of requirements notions. Our goal here is in the ontological classification and conceptual clarification of different requirement kinds.

\subsection{Requirements Specification and Modeling}
\label{sec:art_specificaiton_and_modeling}

In RE, great efforts have been placed to designing effective language for capturing requirements, resulting in various kinds of requirement specification and modeling techniques. We classify these techniques according to their formalities (informal, semi-formal and formal), and show the classification result in Table \ref{tab:classification}. There are two points to be noted. First, we take the viewpoint that modeling is broader than specification: a specification describes the behavior of software system, which is a particular type of model -- behavior model. Second, writing guidelines by themselves are not languages, but empirical rules that help people to use languages. So, we do not classify them into any of the formality categories.

\begin{table}[!htbp]
  \centering
  \small
  \vspace{-0.2cm}
  \begin{threeparttable}
  \caption {A classification of requirements languages}
  \label{tab:classification}
  \setlength\tabcolsep{2pt}
  \begin{tabular}{|c|c|c|l|}
  \hline
  & \textbf{Kind} &  & \textbf{Examples} \\ \hline
  -- & Writing Guidelines & -- & Wiegers et al.~\cite{wiegers_software_2013}\\ \hline
  \multirow{6}{*}{{\rotatebox[origin=c]{90}{\emph{\textbf{Spec}}}}} & Natural Language & \emph{I} & English \\
  & RS Template & \emph{S} & IEEE Std 830~\cite{committee_ieee_1998}, Volere~\cite{robertson_mastering_2012} \\
  & Structured Language & \emph{S} & EARS~\cite{mavin_easy_2009}, Planguage~\cite{gilb_competitive_2005}\\
  & Controlled NL & \emph{S} & ACE~\cite{fuchs_attempto_2006}, Gervasi et al.~\cite{gervasi_reasoning_2005} \\
  & Formal Language & \emph{F} & RML~\cite{greenspan_capturing_1982}, SCR~\cite{heitmeyer_automated_1996}, VDM~\cite{fitzgerald_validated_2005}\\ \hline
  \multirow{3}{*}{{\rotatebox[origin=c]{90}{\emph{\textbf{Mdl}}}}} & Structural Analysis & \emph{S} & $SADT^{TM}$~\cite{ross_structured_1977-1}\\
  & Object Orientation & \emph{S} & UML~\cite{rumbaugh_unified_2004}\\
  & Goal Orientation & \emph{S} & KAOS~\cite{dardenne_goal-directed_1993}, \textit{i}*~\cite{yu_modelling_2011}\\
  \hline
  \end{tabular}
  \begin{tablenotes}
      \item[1] \emph{Spec}: specification; \emph{Mdl}: modeling; \emph{I}: informal; \emph{S}: semi-formal; \emph{F}: formal.
    \end{tablenotes}
  \end{threeparttable}
  \vspace{-0.2cm}
\end{table}


\vspace{6pt}
\noindent \textbf{Requirements Specification Languages}. A common way to express requirements is to use natural language (NL). However, there is much evidence that NL requirements are inherently vague, ambiguous and incomplete~\cite{li_engineering_2016}. 

Requirements specification (RS) templates such as the IEEE Std 830-1998~\cite{committee_ieee_1998} and the Volere template~\cite{robertson_mastering_2012} represent the most basic types of tool for RE. RS templates are useful in classifying and documenting individual requirements, but they offer very limited support for requirements management (e.g., FRs and NFRs are documented separately, the interrelations between them are missing).

To help people write better requirements, writing guidelines have been suggested~\cite{wiegers_software_2013}. These approaches usually use a set of properties of good requirements as criteria, and provide a set of operational guidelines (e.g., avoiding the use of ambiguous words, looking for the ``else'' statement for an ``if'' statement). These techniques are usually informal, and lack a systematic methodology and tool support.


Structured languages (e.g., EARS~\cite{mavin_easy_2009}, Planguage~\cite{gilb_competitive_2005}), have been proposed based on practical experiences, intending to reduce or eliminate certain kinds of requirements issues (e.g., ambiguity and vagueness). There is much evidence that these approaches are effective on their intended use~\cite{mavin_easy_2009}\cite{jacobs_introducing_1999}. However, they are designed exclusively for only FRs or NFRs, and are not for both of them.

Formal languages (e.g., KAOS~\cite{dardenne_goal-directed_1993}) have been advocated because they have a clear syntax and semantics, and promise sophisticated analysis such as ambiguity detection and inconsistency check. Nevertheless, they suffer from major shortcomings~\cite{van_lamsweerde_requirements_2009}: (1) they require high expertise and hence are not really accessible to practitioners or customers; (2) they mainly focus on functional aspects, and leave out important non-functional ones.

Controlled Natural Languages (e.g., ACE~\cite{fuchs_attempto_2006}) combine the advantages of natural and formal languages: (1) practically accessible to engineers and customers; (2) can be mapped to formal language(s) for certain kinds of analysis. However, these approaches do not support refining stakeholder requirements, instead, they assume requirements elicited from stakeholders are in enough detail and can be directly specified, which is not the case in practice.

\vspace{6pt}
\noindent \textbf{Requirements Modeling Languages}. Structured Analysis and Design Technique ($SADT^{TM}$)~\cite{ross_structured_1977-1} is probably the first graphical language used for modeling and communicating requirements, and has served as the starting point of other structured analysis techniques, e.g., the popular data flow diagrams~\cite{demarco_structured_1979}.


Use case, a UML concept that represents the externally visible functionalities of the system-to-be, has been widely used for representing requirements in practice. However, use cases capture only part of the requirements (i.e., FRs), and leave out non-functional ones (e.g., user interface requirements, data requirements, quality requirements)~\cite{cockburn_writing_1999}.

van Lamsweerde and his colleagues have used the concept ``\textit{goal}'', which represents the objective a system under consideration should achieve, to capture requirements, and accordingly proposed KAOS~\cite{dardenne_goal-directed_1993}. At the same period, Mylopoulos et al.~\cite{mylopoulos_representing_1992} have proposed the NFR framework (NFR-F), which uses ``\textit{softgoals}'' (goals without a clear-cut criteria for success) to capture non-functional requirements. These two frameworks pioneered in promoting goal-oriented requirements engineering (GORE).

GORE is advocated for multiple reasons~\cite{van_lamsweerde_goal-oriented_2001}: (1) goals drive the elaboration of requirements and justify requirements; (2) goal models provide a natural mechanism for structuring requirements (AND/OR) and allow reasoning about alternatives; and (3) goal-oriented techniques treat NFRs in depth~\cite{chung_non-functional_2009}. However, goal oriented techniques also have some deficiencies (at the language level)~\cite{li_desiree_2016}. First, they lack a unified language for representing both FRs and NFRs, except natural language that is error prone. Second, they treat individual requirements as propositions (wholes), hence missing important interrelations between requirements (e.g., the dependency relation between qualities and functions).

\subsection{Requirements Transformation}
\label{sec:art_trans}


The early $SADT^{TM}$ proposal~\cite{ross_structured_1977-1} has already suggested useful structural decomposition mechanisms for decomposing a software system into activities and data. In object-oriented development (OOD), the decomposition of a system is based on objects~\cite{booch_object-oriented_1986}. However, both of these approaches have limited scope~\cite{van_lamsweerde_building_2001}: they focus on the software system alone, and do not consider its environment. Moreover, they do not support the capture of NFRs.



With the proposal of ``\textit{goals}'' in the '90s, GORE has been playing a key role in tackling the RE problem. Goal-oriented techniques, such as KAOS~\cite{dardenne_goal-directed_1993}, NFR-F~\cite{chung_non-functional_2000}, \textit{i}*~\cite{yu_modelling_2011}, Tropos~\cite{bresciani_tropos:_2004}, and Techne~\cite{jureta_techne:_2010}, capture stakeholder requirements as goals, use AND/OR refinement to refine high-level strategic goals into low-level operational goals, and use operationalization to operationalize low-level goals as tasks (aka functions). Some of the approaches have used formal languages to formalize requirements specifications (e.g., KAOS, Formal Tropos), enabling certain automatic requirements analysis. However, these existing goal modeling techniques have some common deficiencies: (1) they do not support incrementally improving requirements quality and going from informal to formal; (2) they do not support weakening of requirements.

Problem frames offer two forms of transformation, \emph{decomposition} and \emph{reduction}~\cite{cox_roadmap_2005}, for deriving specifications from stakeholder requirements. \emph{Problem decomposition} allows the transformation of a complex problem into smaller and simpler ones; \emph{Problem reduction} allows to simplify the context of a problem by removing some ($\geq1$) of the domains (the context of a problem is decomposed into many domains) and re-express the requirement using the phenomena in the remaining domains~\cite{cox_roadmap_2005}. The problem with the problem frames approach is that it does not make the distincttion between functional and non-functional requirements, which is widely accepted in the RE field.

\subsection{Empirical evaluation}
\label{sec:art_eval}
In RE, many empirical evaluations have been conducted to assess the utility of some languages or methods, but mainly on their expressiveness and effectiveness~\cite{estrada_empirical_2006}\cite{horkoff_evaluating_2014}.



Al-Subaie et al.~\cite{al-subaie_evaluating_2006} have used a realistic case study to evaluate KAOS and its supporting tool, Objectiver. They reported that KAOS is helpful in detecting ambiguity and capture traceability. However, they also pointed out that the formalism of KAOS is only applicable to goals that are in enough detail and can be directly formalized. 

Matulevicius et al.~\cite{matulevicius_comparing_2007} is quite relevant. In their evaluation, the authors have compared \emph{i}* with KAOS. Beside the quality of languages themselves, they also compared the models generated by using the two framework with regarding to a set of qualitative properties in the semiotic quality framework~\cite{krogstie_semiotic_2002}. Their findings indicate a higher quality of the KAOS language (not significant), but a higher quality of the \emph{i}* models (the participants are not required to write formal specifications with KAOS). They also found that the goal models produced by both frameworks are evaluated low at several aspects, including verifiability, completeness and ambiguity.

These evaluations show that stakeholder requirements initially captured in goal models are of low quality and error prone. The quality of such requirements models needs to be improved, no matter the models will be formalized or not later. That is, techniques for improving the quality of requirements captured in traditional goal models, and incrementally transforming stakeholder requirements to formal specifications are needed. 
\section{Conclusion and Future Work}
\label{cha:conclusion}
In this paper, we presented \emph{Desiree}, a requirements calculus for incrementally transforming stakeholder requirements into an eligible requirements specification, based on a systematic revision of the requirements concepts and operators introduced in our previous work~\cite{li_stakeholder_2015}\cite{li_non-functional_2014}. We formalized the semantics of the requirements concepts and operators, and developed a prototype tool in support of the entire \emph{Desiree} framework.

There are several directions open for our future research. We briefly discuss two most interesting ones in this section.

\textbf{Slot mining}. Our framework currently does not have a built-in set of slots, and may result in different outputs when used by different users. For example, when specifying the relation between ``students'' and ``clinical class'', one may use ``belong to'' while others could use ``associated with''. An interesting idea is to elicit a set of frequently used slots through statistic analysis over corpora.

\textbf{Requirements management}. Our description-based language is able to capture a rich set of interrelations between requirements, functional and non-functional. A promising research direction is to systematically and automatically detect the impact when changing a requirement. It will be very interesting to see how a requirements knowledge base evolves with changing requirements, a major topic in Software Engineering for the next decade.




\section*{Acknowledgment}
This research has been funded by the ERC advanced grant 267856 ``Lucretius: Foundations for Software Evolution'', unfolding during the period of April 2011 - March 2016. It has also been supported by the Key Project of National Natural Science Foundation of China (no. 61432020).

\bibliographystyle{IEEEtran}
\bibliography{desiree}

\end{document}